\documentclass[12pt]{article}

\usepackage[reqno]{amsmath}
\usepackage{mathrsfs}
\usepackage{amssymb}

\usepackage{bbm}
\usepackage{epsfig}
\usepackage{array}
\usepackage{float}
\usepackage{color}
\usepackage{graphicx}

\usepackage{a4}
\usepackage{a4wide}
\usepackage{wasysym}
\usepackage{url}
\usepackage{color}

\usepackage[colorlinks=true,linkcolor=red,citecolor=green,filecolor=magenta,urlcolor=cyan,bookmarks=true,bookmarksopen=true]{hyperref}
\usepackage[all]{hypcap}
\usepackage{mciteplus}

\def\gs{\mathrel{
   \rlap{\raise 0.511ex \hbox{$>$}}{\lower 0.511ex \hbox{$\sim$}}}}
\def\ls{\mathrel{
   \rlap{\raise 0.511ex \hbox{$<$}}{\lower 0.511ex \hbox{$\sim$}}}}

\newcommand{\be}{\begin{eqnarray}}
\newcommand{\ee}{\end{eqnarray}}

\newcommand{\eps}{\mbox{$\epsilon$}}

\def\be{\begin{equation}}
\def\ee{\end{equation}}
\newcommand{\ba}{\begin{array}{c}}
\newcommand{\baz}{\begin{array}{cc}}
\newcommand{\bad}{\begin{array}{ccc}}
\newcommand{\bav}{\begin{array}{cccc}}
\newcommand{\baf}{\begin{array}{ccccc}}
\newcommand{\bena}{\begin{eqnarray}}
\newcommand{\eena}{\end{eqnarray}}
\newcommand{\bea}{\begin{equation} \begin{array}{c}}
\newcommand{\eea}{ \end{array} \end{equation}}
\newcommand{\ea}{\end{array}}

\def\eps{\epsilon}

\begin{document}

\begin{titlepage}
\title{\vspace*{-2.0cm}
\bf\Large
Soft $L_e-L_\mu-L_\tau$ flavour symmetry breaking and sterile neutrino keV Dark Matter
\\[5mm]\ }

\author{
Manfred Lindner$^a$\thanks{email: \tt lindner@mpi-hd.mpg.de}~~,~~
Alexander Merle$^{ab}$\thanks{email: \tt amerle@kth.se}~~, and~~
Viviana Niro$^a$\thanks{email: \tt viviana$.$niro@mpi-hd.mpg.de}
\\ \\
$^a${\normalsize \it Max-Planck-Institut f\"ur Kernphysik,}\\
{\normalsize \it Postfach 10 39 80, 69029 Heidelberg, Germany}\\
\\
$^b${\normalsize \it Department of Theoretical Physics, School of Engineering Sciences,}\\
{\normalsize \it Royal Institute of Technology (KTH), AlbaNova University Center,}\\
{\normalsize \it Roslagstullsbacken 21, 106 91 Stockholm, Sweden}
}

\date{\today}
\maketitle
\thispagestyle{empty}

\begin{abstract}
\noindent
We discuss how a $L_e-L_\mu-L_\tau$ flavour symmetry that is softly broken leads to keV sterile neutrinos, which are a prime candidate for Warm Dark Matter. This is to our knowledge the first model where flavour symmetries are applied simultaneously to active and sterile neutrinos explaining at the same time active neutrino properties and this peculiar Dark Matter scenario. The essential point is that different scales of the symmetry breaking and the symmetry preserving entries in the mass matrix lead to one right-handed neutrino which is nearly massless compared to the other two. Furthermore, we naturally predict vanishing $\theta_{13}$ and maximal $\theta_{23}$, while the correct value of $\theta_{12}$ must come from the mixing of the charged leptons. We can furthermore predict an exact mass spectrum for the light neutrinos, which will be testable in the very near future.
\end{abstract}

\end{titlepage}

\section{\label{sec:intro} Introduction}

The nature of Dark Matter (DM) is a longstanding problem of particle physics and astrophysics, see 
e.g.~\cite{Zwicky,D'Amico:2009df,Bertone:2004pz}. New elementary particles with suitable properties are predicted by various extensions of the SM and some of them are excellent candidates to make up the DM in the Universe. Any candidate does not only have to yield the right DM abundance, $\Omega_{\rm DM} h^2 \simeq 0.1$~\cite{Komatsu:2010fb}, but it also has to be consistent with bounds coming from structure formation provided by the Sloan Digital Sky Survey (SDSS)~\cite{SDSS} and by the 2dF Galaxy Redshift Survey (2dF-GRS)~\cite{2dF}; see also the analysis of the so-called Lyman-$\alpha$ Forest~\cite{Weinberg:2003eg}. These bounds rule out hot (relativistic) DM as the dominant component, which is why ordinary (active) neutrinos are excluded as Dark Matter candidate~\cite{Komatsu:2010fb}. In addition active neutrinos are too light to explain the required DM abundance. A good alternative is that DM is cold (non-relativistic), which fits very well to the cosmological standard model ($\Lambda$CDM), 
which is in perfect agreement to observations~\cite{Komatsu:2010fb}.

A less explored possibility is so-called `Warm Dark Matter', for which a single light sterile neutrino would be a natural candidate~\cite{Dolgov:2000ew}. 
The advantage of such a scenario would be that right-handed neutrinos most likely exist anyhow in order to accommodate neutrino masses and one needs only some extra feature which explains why one sterile state is light in the multi keV range. Model-independent analyses of the data even seem to point into the direction of keV-masses for DM~\cite{deVega:2009ku}. Such a scenario with three active plus one sterile light neutrino states fits also very well to the number of light neutrino species obtained from recent analyses of Big Bang Nucleosynthesis data~\cite{Izotov:2010ca}. A simple framework is to assume a certain parameter choice for the Standard Model (SM) extended by three right-handed neutrinos, which is called `$\nu$SM'~\cite{Asaka:2005an}. This setting has been shown to be in agreement with cosmology~\cite{Asaka:2006nq,Asaka:2006rw,Roy:2010xq,Boyarsky:2005us,Boyarsky:2006zi,Boyarsky:2008mt} and could also lead to detectable signals in laboratory experiments~\cite{Bezrukov:2006cy,Liao:2010yx,Li:2010vy}.

Another setting, which depends less on physics at even higher scale has recently been discussed by Bezrukov, Hettmansperger, and Lindner~\cite{Bezrukov:2009th}. They studied right-handed neutrinos which couple to gauge extensions which naturally leads to the right DM abundance. In this context, it has been shown that keV sterile neutrino DM can be brought into agreement with current bounds in a type~II seesaw framework, while type~I scenarios are practically excluded. Apart from invoking a low scale seesaw mechanism, however, this paper did not 
provide an explanation why one sterile neutrino should have a keV-ish mass, while all the others are very heavy. One attempt to explain such a pattern was recently put forward by Kusenko, Takahashi, and Yanagida~\cite{Kusenko:2010ik} in a scenario with extra dimensions leading to a `split see-saw'. Here, we attempt to explain the required mass pattern by using $L_e-L_\mu-L_\tau$ flavour symmetry 
in order to obtain keV sterile neutrino Dark Matter. Note that this would imply that the same flavour symmetry, which explain the masses and mixings of active neutrinos, would also affect the pattern of sterile neutrinos and explain in this way a natural DM scenario. 
Our study can be applied to the model of Ref.~\cite{Bezrukov:2009th}, but also to any other model that present a keV sterile neutrino 
as DM candidate.

Our key idea is the following: Using $L_e-L_\mu-L_\tau$, we can already explain hierarchical neutrino spectra with one neutrino out of three being massless. Soft breaking of this symmetry brings in a completely new scale, much lower than the original one, which will determine the size of the induced mass of the neutrino that was massless before. Such a setup could very well point to scenarios with one sterile neutrino with a keV-mass, while the other two are heavier, which could be promising for cosmology as well as for laboratory detections~\cite{Liao:2010yx}.
The key point is that the same flavour symmetry can be responsible for similar structures in the heavy and in the light neutrino sector. In fact, all the cases studied by us predict a strong (inverted) hierarchy for the light neutrinos, which can be tested in the near future.

Furthermore, the symmetry predicts bimaximal mixing coming from the neutrino sector. This involves a maximal mixing angle $\theta_{12}$, which is ruled out experimentally at more than 6$\sigma$. However, since there is also mixing coming from the charged leptons, it is possible to alleviate or even solve this problem.

Our approach will be pragmatical in the sense that we simply consider an example of mixing matrix in the charged lepton sector that can lead to allowed values of neutrino mixing angles. We furthermore argue that it is possible to find charged lepton mass matrices which yield this mixing without contradicting $L_e-L_\mu-L_\tau$, in the sense that the largest elements of the mass matrix preserve the symmetry. Of course, one could try to go into some more detail here by having a closer look at the flavon potential and its breaking chain, in order to enforce this particular structure for the charged lepton mass matrix. This would, however, lead beyond the scope of this paper and would rather distract the reader from the simple statement we want to make, so we leave this point to further studies and focus on the neutrino part instead.

The paper is organized as follows: In Sec.~\ref{sec:TBM} we generally describe what mixing we need in the charged lepton sector in order to translate a bimaximal neutrino mixing into a Pontecorvo-Maki-Nakagawa-Sakata (PMNS) matrix that is compatible with the experimental values. We then discuss our general framework and three different models in Sec.~\ref{sec:model}, before finally concluding in Sec.~\ref{sec:conclusions}.

\section{\label{sec:TBM} The PMNS matrix}

Before we start with a discussion of the models under consideration, we shortly comment on the mixing that is predicted, and on how to bring it to agreement with experimental data.

The PMNS matrix $\mathcal{U}$ is often parametrized as:
\begin{equation}
\mathcal{U}=
\begin{pmatrix}
c_{12}c_{13} & s_{12} c_{13} & s_{13}\\
-s_{12} c_{23} - c_{12} s_{23} s_{13}\,e^{i \delta_{\rm CP}} & 
c_{12} c_{23} - s_{12} s_{23} s_{13}\,e^{i \delta_{\rm CP}} & 
s_{23} c_{13} \,e^{i\delta_{\rm CP}}\\
s_{12} s_{23} - c_{12} c_{23} s_{13}\,e^{i \delta_{\rm CP}} & 
- c_{12} s_{23} - s_{12} c_{23} s_{13}\,e^{i \delta_{\rm CP}} &
c_{23} c_{13}\,e^{i \delta_{\rm CP}}
\end{pmatrix}\,\cdot \textrm{diag}(1,\,e^{i\alpha},\,e^{i\beta}),
\label{eq:Umatrix}
\end{equation}
where $c_{ij} \equiv \cos \theta_{ij}$ and $s_{ij} \equiv \sin \theta_{ij}$. The parameter $\delta_{\rm CP}$ is the Dirac CP-violating phase, while $\alpha$ and $\beta$ are the two Majorana phases, absent in the case of Dirac neutrinos. The PMNS matrix is given by $\mathcal{U}=\mathcal{U}^\dagger_L \mathcal{U}_\nu$, where $\mathcal{U}_\nu$ is the matrix that diagonalizes the neutrino mass matrix, while $\mathcal{U}_L$ contributes to the diagonalization of the charged lepton mass matrix.

The PMNS mixing matrix should provide values of the neutrino mixing angles 
compatible with the $3\sigma$ intervals allowed by the solar, atmospheric, reactor (KamLAND and CHOOZ), and accelerator (K2K and MINOS) experiments~\cite{Schwetz:2008er}:
\bena
0.25 \le & \sin^{2}\theta_{12} & \le 0.37\,,\nonumber\\
0.36 \le & \sin^{2}\theta_{23} & \le 0.67\,,\\
&\sin^{2}\theta_{13} & \le 0.056\,.\nonumber
\eena
The allowed best-fit values and $1\sigma$ errors of the mass square differences are given by $\Delta m^{2}_{21} = 7.65^{+0.23}_{-0.20}\times 10^{-5}\,{\rm eV^{2}}$ and $|\Delta m^{2}_{31}| = 2.40^{+0.12}_{-0.11}\times 10^{-3}\,{\rm eV^{2}}$. The sign of $\Delta m^2_{31}$ is still unknown. It can be positive for normal mass ordering ($m_1<m_2<m_3$) or negative for inverted mass ordering ($m_3<m_1<m_2$). No experimental information on the value of $\delta_{\textrm{CP}}$ is present at the moment.

A form of the mixing matrix which is frequently used and compatible with all current data is the so-called tri-bimaximal (TBM) mixing:
\begin{equation}
\mathcal{U}_{\rm TBM}=
\begin{pmatrix}
\sqrt{\frac{2}{3}} & \frac{1}{\sqrt{3}} & 0\\
-\frac{1}{\sqrt{6}} & \frac{1}{\sqrt{3}} & \frac{1}{\sqrt{2}} \\
\frac{1}{\sqrt{6}} & -\frac{1}{\sqrt{3}} & \frac{1}{\sqrt{2}}
\end{pmatrix}.
\label{eq:TBM}
\end{equation}
Details on the requirements for mass matrices in order to yield TBM can be found in Ref.~\cite{Plentinger:2005kx}. 
Exact $L_e-L_\mu-L_\tau$ symmetry generically predicts a bimaximal form for the neutrino mixing matrix~\cite{Petcov:2004rk},
\begin{equation}
 \mathcal{U}_\nu=
\begin{pmatrix}
\frac{1}{\sqrt{2}} & \frac{1}{\sqrt{2}} & 0\\
-\frac{1}{2} & \frac{1}{2} & \frac{1}{\sqrt{2}} \\
\frac{1}{2} & -\frac{1}{2} & \frac{1}{\sqrt{2}}
\end{pmatrix}\,.
 \label{eq:bimax}
\end{equation}
The PMNS matrix can deviate from the bimaximal structure reported above, depending on the actual form of $\mathcal{U}_L$. We can parametrize the matrix $\mathcal{U}_L$ as a function of three mixing angles, in analogy to Eq.~\eqref{eq:Umatrix}. For definiteness, we will focus on CP conservation and we will define $\lambda_{ij}\equiv \sin \theta^\prime_{ij}$. Considering a hierarchical relation between $\lambda_{ij}$, of the type $\lambda_{12}=\lambda$, $\lambda_{13}\simeq\lambda^3$, $\lambda_{23}\simeq \lambda^2$, with $\lambda\simeq 0.20$ being the parameter that describes the deviation of $\theta_{12}$ from $\pi/4$~\cite{Frampton:2004ud}, we find
\begin{equation}
 \mathcal{U}_L=
\begin{pmatrix}
1-\lambda^2/2 & \lambda & \lambda^3\\
-\lambda & 1-\lambda^2/2 & \lambda^2\\
\lambda^3 & -\lambda^2 & 1
\end{pmatrix}+\mathcal{O}(\lambda^4)\,.
 \label{eq:UL}
\end{equation}
It was shown in Ref.~\cite{Frampton:2004ud} that this form of $\mathcal{U}_L$, combined with the 
bimaximal matrix of Eq.~\eqref{eq:bimax}, could lead to $\mathcal{U}_{\rm PMNS}$ compatible with the experimental values:
\bena
\tan^2\theta_{12}\simeq 1-2\sqrt{2}\lambda+4\lambda^4-2\sqrt{2}\lambda^3
&\rightarrow&\theta_{12}\simeq 33.4^{\circ}\,,
\label{eq:theta12}\\
|U_{e3}|\simeq \frac{\lambda}{\sqrt{2}}
&\rightarrow&\theta_{13}\simeq 8^{\circ}\,,
\label{eq:theta13}\\
\sin^2 2\theta_{23}\simeq 1-4\lambda^4 
&\rightarrow&\theta_{23}\simeq45^{\circ}\,.
\label{eq:theta23}
\eena
The matrix $\mathcal{U}_L$ is associated to the diagonalization of the matrix $\mathcal{M}_l\mathcal{M}^\dagger_l$, 
with $\mathcal{M}_l$ being the charged leptons mass matrix. The expression of $U_L$ reported in Eq.~\eqref{eq:UL} 
will diagonalize the matrix
\begin{equation}
 \mathcal{M}_l \mathcal{M}^\dagger_l \simeq
\begin{pmatrix}
m^2_e+m^2_\mu\lambda^2 & m^2_\mu\lambda & 0\\
m^2_\mu \lambda & m^2_\mu & 0\\
0 & 0 & m^2_\tau
\end{pmatrix}\,,
\label{eq:Mlsymmetry}
\end{equation}
where we have neglected terms of order $\lambda^3$. In the next section we will show under which assumptions the charged lepton mass matrix $\mathcal{M}_l \mathcal{M}^\dagger_l $ can be in agreement with (softly broken) $L_e-L_\mu-L_\tau$ symmetry.

\section{\label{sec:model} The $L_e-L_\mu-L_\tau$ flavour symmetry}

Let us define the symmetry $\mathcal{F}\equiv L_e-L_\mu-L_\tau$. We extend the model~\cite{Lavoura:2000ci} by a Higgs triplet~$\Delta$, in order to accommodate a type II seesaw, which is required in the context of the model presented in Ref.~\cite{Bezrukov:2009th}. The particle content and the charge assignments of our model are given by:
\begin{center} 
\begin{tabular}{|c|c|c|c|c|c|c|c|c|c|c|c|}\hline
  & $L_{e L}$ & $L_{\mu L}$ & $L_{\tau L}$ & 
$e_R$ & $\mu_R$ & $\tau_R$ &
$N_{1 R}$ & $N_{2 R}$ & $N_{3 R}$ & 
$\phi$ & $\Delta$ \\\hline
 $\mathcal{F}$ & $1$ & $-1$ & $-1$ &
$1$ & $-1$ & $-1$ &
$1$ & $-1$ & $-1$ &
$0$ & $0$   \\\hline
 \end{tabular}
\end{center}
where $L_{\alpha L}=(\nu_{\alpha L}, \alpha_L)^T$, and $\alpha=e, \mu, \tau$. 
The scalars are defined as:
\bena
\phi=
\begin{pmatrix}
\phi^+ \\
\phi^0
\end{pmatrix}\,,& & 
\Delta=
\begin{pmatrix}
\Delta^+/\sqrt{2} & \Delta^{++}\\
\Delta^0 & -\Delta^+/\sqrt{2}
\end{pmatrix}\,.\nonumber
\eena

The right-handed neutrino fields are gauge singlets and we can thus write down the following symmetry conserving Majorana mass term for the right-handed neutrinos:
\begin{equation}
\mathcal{L}_{\rm mass}=-M^{12}_R \, \, \overline{(N_{1 R})^C} \, N_{2 R} -
M^{13}_R\, \, \overline{(N_{1 R})^C} \, N_{3 R} + h.c.
\label{eq:nu_mass_right}
\end{equation}

The Dirac mass term, which links the left- and right-handed fields, is given by:
\bena
\mathcal{L}_{\rm mass}&=&-Y^{e 1}_D \, \, \overline{L_{e L}}\,\tilde{\phi}\,N_{1 R} - 
Y^{\mu 2}_D \, \, \overline{L_{\mu L}}\,\tilde{\phi}\,N_{2 R} -
Y^{\mu 3}_D \, \, \overline{L_{\mu L}}\,\tilde{\phi}\,N_{3 R} -\nonumber\\
& & -
Y^{\tau 2}_D \, \, \overline{L_{\tau L}}\,\tilde{\phi}\,N_{2 R}-
Y^{\tau 3}_D \, \, \overline{L_{\tau L}}\,\tilde{\phi}\,N_{3 R} + h.c.,
\label{eq:nu_mass_dirac}
\eena
with $\tilde{\phi}=i \sigma_2 \phi^{*}$. Using the triplet scalar, we can also have a Majorana mass term for the left-handed neutrinos:
\begin{equation}
\mathcal{L}_{\rm mass}=-Y^{e \mu}_L \, \, 
\overline{(L_{e L})^C}\,(i \sigma_2 \Delta)\,L_{\mu L} - 
Y^{e \tau}_L \, \, \overline{(L_{e L})^C}\,(i \sigma_2 \Delta)\,L_{\tau L} + h.c.
\label{eq:nu_mass_left}
\end{equation}

In total, considering Eqs.~\eqref{eq:nu_mass_right}, \eqref{eq:nu_mass_dirac}, and~\eqref{eq:nu_mass_left}, the neutrino mass term can be written as:
\begin{equation}
\mathcal{L}_{\rm mass}=-\frac{1}{2} 
\overline{\Psi^C} \mathcal{M}_\nu \Psi +h.c.\,,
\label{eq:nu_mass_tot}
\end{equation}
with $\Psi \equiv \left((\nu_{e L})^C,(\nu_{\mu L})^C,(\nu_{\tau L})^C,N_{1 R},N_{2 R},N_{3 R}\right)^T$ and 
\begin{equation}
\mathcal{M}_\nu=
\begin{pmatrix}
\begin{array}{c|c}
\begin{matrix}
0 & m^{e \mu}_L & m^{e \tau}_L \\
m^{e \mu}_L & 0 & 0 \\
m^{e \tau}_L & 0 & 0 
\end{matrix}
&
\begin{matrix}
m^{e 1}_D & 0 & 0 \\
0 & m^{\mu 2}_D & m^{\mu 3}_D \\
0 & m^{\tau 2}_D & m^{\tau 3}_D
\end{matrix}\\\hline
\begin{matrix}
m^{e 1}_D & 0 & 0 \\
0 & m^{\mu 2}_D & m^{\tau 2}_D \\
0 & m^{\mu 3}_D & m^{\tau 3}_D
\end{matrix}
&
\begin{matrix}
0 & M^{12}_R & M^{13}_R \\
M^{12}_R & 0 & 0 \\
M^{13}_R & 0 & 0
\end{matrix}
\end{array}
\end{pmatrix},
\end{equation}
where we have defined $m^{\alpha i}_D = v_\phi Y^{\alpha i}_D$ and $m^{\alpha \beta}_L = v_\Delta Y^{\alpha \beta}_L$. $v_\phi$ and $v_\Delta$ are the vacuum expectation values (VEVs) of the scalar doublet $\phi$ and the triplet $\Delta$.\footnote{Note that the VEV of the latter is constrained by corrections to the $\rho$ parameter to be at most about $1$~GeV.} We can diagonalize the symmetric mass matrix $\mathcal{M}_\nu$ with the help of an orthogonal matrix $\mathcal{O}$. We can study three different regimes for the parameters of the mass matrix:
\begin{itemize}
 \item $m^{\alpha i}_D \ll m^{\alpha \beta}_L \ll M^{i j}_R$ (separation scenario),
 \item $m^{\alpha \beta}_L \ll m^{\alpha i}_D \ll M^{i j}_R$ (type II see-saw scenario),
 \item $m^{\alpha \beta}_L \sim m^{\alpha i}_D \ll M^{i j}_R$ (hybrid scenario).
\end{itemize}
For simplicity, throughout our discussion, we will assume the parameters $m^{\alpha i}_D$, $m^{\alpha \beta}_L$, and $M^{i j}_R$ to be real. We will show in the following how soft-breaking terms could alter the explicit 
form of the eigenvalues associated with $\mathcal{M}_l$ and $\mathcal{M}_\nu$.

The mass term for the charged leptons will also experience restrictions by the $\mathcal{F}$-symmetry and is given by:
\bena
\mathcal{L}_{\rm mass}&=&-Y^{e e}_D \, \, \overline{L_{e L}}\,\phi\,e_{R} - 
Y^{\mu \mu}_D \, \, \overline{L_{\mu L}}\,\phi\,\mu_{R} -
Y^{\mu \tau}_D \, \, \overline{L_{\mu L}}\,\phi\,\tau_{R} -\nonumber\\
& & -
Y^{\tau \mu}_D \, \, \overline{L_{\tau L}}\,\phi\,\mu_{R}-
Y^{\tau \tau}_D \, \, \overline{L_{\tau L}}\,\phi\,\tau_{R} + h.c.,
\label{eq:charged_mass}
\eena
which can be rewritten in matrix form:

\begin{equation}
\mathcal{L}_{\rm mass}=-
\left(\overline{e_L},\overline{\mu_L},\overline{\tau_L} \right) \mathcal{M}_l 
\begin{pmatrix}
e_R\\
\mu_R\\
\tau_R
\end{pmatrix}
+ h.c.,
\label{eq:nu_mass_tot2}
\end{equation}
with the mass matrix $\mathcal{M}_l$ given by
\begin{equation}
\mathcal{M}_l=
\begin{pmatrix}
m^{e e}_D & 0 & 0 \\
0 & m^{\mu \mu}_D & m^{\mu \tau}_D \\
0 & m^{\tau \mu}_D & m^{\tau \tau}_D
\end{pmatrix}.
\label{eq:ML_form}
\end{equation}
The charged lepton mass matrix $\mathcal{M}_l$ is diagonalized by a bi-unitary transformation,
\begin{equation}
\mathcal{U}^\dagger_L\,\mathcal{M}_l\,\mathcal{U}_R=
\mathcal{M}^{\rm diag}_l=
\begin{pmatrix}
m_e &  0    &  0     \\
 0  & m_\mu &  0     \\
 0  &  0    & m_\tau
\end{pmatrix}\,,
\end{equation}
where $m_e$, $m_\mu$, $m_\tau$ are the electron, the muon, and the tau masses, respectively. 
We can then assume the presence of terms that break the $L_e-L_\mu-L_\tau$ symmetry softly 
through $|\Delta \mathcal{F}|=2$ Dirac mass terms:
\bena
\mathcal{L}_{\rm soft}&=&-Y^{e \mu}_D \, \, \overline{L_{e L}}\,\phi\,\mu_{R} - 
Y^{e \tau}_D \, \, \overline{L_{e L}}\,\phi\,\tau_{R} -
Y^{\mu e}_D \, \, \overline{L_{\mu L}}\,\phi\,e_{R} -\nonumber\\
& & -
Y^{\tau e}_D \, \, \overline{L_{\tau L}}\,\phi\,e_{R} + h.c.,
\eena
where we can define $s^{\alpha \beta}_D=v_\phi Y^{\alpha \beta}_D$ and 
$\alpha,\beta=e,\mu,\tau$. 
Considering, for simplicity, a symmetric form of the charged lepton mass matrix 
($s^{\tau e}_D=s^{e \tau}_D$, $s^{\mu e}_D=s^{e \mu}_D$, and $m^{\tau \mu}_D=m^{\mu \tau}_D$), we have:
\begin{equation}
\mathcal{M}_l=
\begin{pmatrix}
m^{e e}_D & s^{e \mu}_D & s^{ e \tau}_D \\
s^{e \mu}_D & m^{\mu \mu}_D & m^{\mu \tau}_D \\
s^{e \tau}_D & m^{\mu \tau}_D & m^{\tau \tau}_D
\end{pmatrix}.
\end{equation}
Under the assumption of small $s^{e \tau}_D$ and $m^{\mu \tau}_D$, we obtain 
\begin{equation}
\mathcal{M}_l \mathcal{M}^\dagger_l=
\begin{pmatrix}
\left(m^{e e}_D\right)^2 + \left(s^{e \mu}_D\right)^2 & s^{e \mu}_D\left(m^{e e}_D+m^{\mu \mu}_D\right) & 0 \\
s^{e \mu}_D\left(m^{e e}_D+m^{\mu \mu}_D\right) & \left(m^{\mu \mu}_D\right)^2 + \left(s^{e \mu}_D\right)^2 & 0 \\
0 & 0 & \left(m^{\tau \tau}_D\right)^2
\end{pmatrix}.
\end{equation}
If we identify $s^{e \mu}_{D}=m^{\mu \mu}_D \lambda$, $m^{e e}_{D}=m_e$, 
$m^{\mu \mu}_{D}=m_\mu$, and $m^{\tau \tau}_{D}=m_\tau$, we get an $\mathcal{M}_l \mathcal{M}^\dagger_l$ similar to Eq.~\eqref{eq:Mlsymmetry}. 
In this case, we would have a charged lepton mixing matrix $\mathcal{U}_L$ like the one in Eq.~\eqref{eq:UL}. We will show in the following discussion that the 
neutrino mixing matrix is compatible with the bimaximal form, even in the 
presence of soft-breaking terms. Therefore, 
the PMNS matrix will lead to the values of the neutrino mixing angles reported in 
Eqs.~\eqref{eq:theta12}, \eqref{eq:theta13}, and~\eqref{eq:theta23}. We assume the problem of $\theta_{12}$ to be solved in this way. The task to investigate the details of a corresponding flavon potential will be left to others, as it would lead away from our main point, which is the prediction for the neutrino sector. We want to stress that any model with $L_e-L_\mu-L_\tau$ symmetry will lead to similar predictions concerning the neutrino mass spectra, no matter how the problem of getting the correct $\theta_{12}$ is solved.

\subsection{Case 1 (separation scenario): $m^{\alpha i}_D \ll m^{\alpha \beta}_L \ll M^{i j}_R$}

Under the assumption that $m^{\alpha i}_D \ll m^{\alpha \beta}_L \ll M^{i j}_R$, 
with $i, j=1, 2, 3$ and $\alpha, \beta = e, \mu, \tau$, we obtain:
\begin{equation}
\mathcal{O}^T\,\mathcal{M}^{(1)}_\nu\,\mathcal{O}=
\mathcal{M}^{\rm (1),diag}_\nu=
\begin{pmatrix}
\lambda_+ & 0 & 0 & 0 & 0 & 0\\
0 & \lambda_- & 0 & 0 & 0 & 0\\
0 & 0 & 0 & 0 & 0 & 0\\
0 & 0 & 0 & \Lambda_+ & 0 & 0\\
0 & 0 & 0 & 0 & \Lambda_- & 0\\
0 & 0 & 0 & 0 & 0 & 0\\
\end{pmatrix},
\end{equation}
with $\lambda_\pm =~\pm \sqrt{(m^{e \mu}_L)^2 + (m^{e \tau}_L)^2}$ and 
$\Lambda_\pm =~\pm \sqrt{(M^{12}_R)^2 + (M^{13}_R)^2}$. 
Since $\lambda_-$ and $\Lambda_-$ are negative eigenvalues of the mass 
matrix $\mathcal{M}_\nu$, we choose to define the neutrino mass basis as 
$\chi \equiv(\nu_{1 R},-i \nu_{2 R}, \nu_{3 R}, \nu_{4 R}, -i \nu_{5 R}, \nu_{6 R})^T$. 
The orthogonal matrix $\mathcal{O}$ is then given by
\begin{equation}
\mathcal{O}=
\begin{pmatrix}
\frac{1}{\sqrt{2}} & \frac{1}{\sqrt{2}} & 0 & 0 & 0 & 0 \\
-\frac{m_{L1}}{\sqrt{2} \sqrt{m_{L1}^2+m_{L2}^2}} & \frac{m_{L1}}{\sqrt{2} \sqrt{m_{L1}^2+m_{L2}^2}} & \frac{m_{L2}}{\sqrt{m_{L1}^2+m_{L2}^2}} & 0 & 0 & 0 \\
   \frac{m_{L2}}{\sqrt{2} \sqrt{m_{L1}^2+m_{L2}^2}}     &    -\frac{m_{L2}}{\sqrt{2} \sqrt{m_{L1}^2+m_{L2}^2}}   &  \frac{m_{L1}}{\sqrt{m_{L1}^2+m_{L2}^2}}&    0     &     0     &  0 \\
   0     &     0     &  0 & \frac{1}{\sqrt{2}} & \frac{1}{\sqrt{2}} & 0\\
   0     &     0     &  0 & -\frac{M_{R 1}}{\sqrt{2} \sqrt{M_{R 1}^2+M_{R 2}^2}} & \frac{M_{R 1}}{\sqrt{2} \sqrt{M_{R 1}^2+M_{R 2}^2}} & \frac{M_{R 2}}{\sqrt{M_{R 1}^2+M_{R 2}^2}} \\
   0     &     0     &  0 &    \frac{M_{R 2}}{\sqrt{2} \sqrt{M_{R 1}^2+M_{R 2}^2}}     &   
-\frac{M_{R 2}}{\sqrt{2} \sqrt{M_{R 1}^2+M_{R 2}^2}}    & \frac{M_{R 1}}{\sqrt{M_{R 1}^2+M_{R 2}^2}}
\end{pmatrix}\,,
\end{equation}
where we have set $m^{e \mu}_L= m_{L1}$, $m^{e \tau}_L=m_{L2}$, $M^{12}_R=M_{R 1}$ and 
$M^{13}_R=M_{R 2}$. The neutrino interaction basis $\Psi$ is related to the neutrino mass basis 
$\chi$ by
\begin{equation}
\Psi=\mathcal{O}^T \chi\,.
\end{equation}
Defining $\tan \theta = \frac{m_{L2}}{m_{L1}}$ and $\tan \psi = \frac{M_{R 2}}{M_{R 1}}$ we obtain:
\begin{equation}
\mathcal{O}=
\begin{pmatrix}
\frac{1}{\sqrt{2}}            & \frac{1}{\sqrt{2}}             & 0 
& 0 & 0 & 0 \\
-\frac{1}{\sqrt{2}}\cos \theta & \frac{1}{\sqrt{2}}\cos \theta & \sin \theta 
& 0 & 0 & 0 \\
\frac{1}{\sqrt{2}}\sin \theta & -\frac{1}{\sqrt{2}}\sin \theta &  \cos \theta 
& 0 & 0 &  0 \\
0 & 0 & 0 & 
\frac{1}{\sqrt{2}}            & \frac{1}{\sqrt{2}}             & 0\\
0 & 0 & 0 & 
-\frac{1}{\sqrt{2}}\cos \psi & \frac{1}{\sqrt{2}}\cos \psi & \sin \psi \\
0 & 0 & 0 &
\frac{1}{\sqrt{2}}\sin \psi & -\frac{1}{\sqrt{2}}\sin \psi &  \cos \psi \\
\end{pmatrix}=
\begin{pmatrix}
\mathcal{U}_\nu & 0\\
0           & \mathcal{W}_\nu
\end{pmatrix}.
\label{eq:bimaximal}
\end{equation}
This form of the light neutrino mixing matrix $\mathcal{U}_\nu$ is exactly bimaximal if $\theta=\frac{\pi}{4}$ (cf.\ Eq.~\eqref{eq:bimax}), which will happen in the limit $m_{L1}=m_{L2}$. To avoid the presence of two massless eigenstates ($m_3=0$ and $M_3=0$), 
we can assume that the $L_e-L_\mu-L_\tau$ flavour symmetry is softly 
broken by $|\Delta \mathcal{F}|=2$ Majorana mass terms:
\bena
\mathcal{L}_{\rm soft}&=&-\frac{1}{2} 
\left(
Y^{ee}_L \, \, \overline{(L_{e L})^C}\,(i \sigma_2 \Delta)\,L_{e L} +
Y^{\mu \mu}_L \, \, \overline{(L_{\mu L})^C}\,(i \sigma_2 \Delta)\,L_{\mu L} +  
Y^{\tau \tau}_L \, \, \overline{(L_{\tau L})^C}\,(i \sigma_2 \Delta)\,L_{\tau L} 
\right)
\nonumber\\
&&-\frac{1}{2} 
\left(S^{11}_R \, \, \overline{(\nu_{1 R})^C} \, \nu_{1 R} + 
S^{22}_R \, \, \overline{(\nu_{2 R})^C} \, \nu_{2 R} + 
S^{33}_R \, \, \overline{(\nu_{3 R})^C} \, \nu_{3 R} \right) + h.c.
\eena
In this case, the following mass matrix results:
\begin{equation}
\mathcal{M}^{(1)}_\nu=
\begin{pmatrix}
\begin{array}{c|c}
\begin{matrix}
s^{e e}_L & m^{e \mu}_L & m^{e \tau}_L \\
m^{e \mu}_L & s^{\mu \mu}_L & 0 \\
m^{e \tau}_L & 0 & s^{\tau \tau}_L 
\end{matrix}
&
\begin{matrix}
m^{e 1}_D & 0 & 0 \\
0 & m^{\mu 2}_D & m^{\mu 3}_D \\
0 & m^{\tau 2}_D & m^{\tau 3}_D
\end{matrix}\\\hline
\begin{matrix}
m^{e 1}_D & 0 & 0 \\
0 & m^{\mu 2}_D & m^{\tau 2}_D \\
0 & m^{\mu 3}_D & m^{\tau 3}_D
\end{matrix}
&
\begin{matrix}
S^{11}_R & M^{12}_R & M^{13}_R \\
M^{12}_R & S^{22}_R & 0 \\
M^{13}_R & 0 & S^{33}_R
\end{matrix}
\end{array}
\end{pmatrix},
\end{equation}
where $s^{\alpha \alpha}_L = v_\Delta Y^{\alpha \alpha}_L$ and 
$\alpha=e, \mu, \tau$. Considering for simplicity 
$s^{\alpha \alpha}_L \simeq s$, $S^{ii}_R \simeq S$, and 
$m^{\alpha i}_D \ll m^{\alpha \beta}_L \ll M^{i j}_R$, we obtain after diagonalization
\begin{equation}
\mathcal{O}^T\,\mathcal{M}^{(1)}_\nu\,\mathcal{O}=
\mathcal{M}^{\rm (1),diag}_\nu=
\begin{pmatrix}
\lambda^\prime_+ & 0 & 0 & 0 & 0 & 0\\
0 & \lambda^\prime_- & 0 & 0 & 0 & 0\\
0 & 0 & \lambda_s & 0 & 0 & 0\\
0 & 0 & 0 & \Lambda^\prime_+ & 0 & 0\\
0 & 0 & 0 & 0 & \Lambda^\prime_- & 0\\
0 & 0 & 0 & 0 & 0 & \Lambda_s\\
\end{pmatrix},
\end{equation}
with $\lambda^\prime_\pm = s \pm \sqrt{(m^{e \mu}_L)^2 + (m^{e \tau}_L)^2}$, $\lambda_s=s$, $\Lambda^\prime_\pm = S \pm \sqrt{(M^{12}_R)^2 + (M^{13}_R)^2}$, and $\Lambda_s = S$. In case of $S \ll M^{12}_R,\, M^{13}_R$ we have $\Lambda^\prime_\pm \simeq \pm \sqrt{(M^{12}_R)^2 + (M^{13}_R)^2}$. 
In this way we could explain the presence of one keV 
sterile neutrino ($S \simeq$ keV) and two heavier sterile neutrinos ($M^{12}_R,\,M^{13}_R \gg S$), as 
required by the working example reported in Ref.\,\cite{Bezrukov:2009th}. For the light neutrinos, we have $m_1=s+b$, $m_2=s-b$, and $m_3=s$, where $b=\sqrt{(m^{e \mu}_L)^2 + (m^{e \tau}_L)^2}$. Note that we need $s<0$ due to the condition $|m_1|<|m_2|$, which leads to physical light neutrino masses of $|m_1|=b-|s|$, $|m_2|=b+|s|$, and $|m_3|=|s|$. The resulting light neutrino mass square differences, in case of $|s| \ll |m^{e \mu,e \tau}_L|$, are given by $\Delta m^2_{21} \simeq 4 (-s) b>0$ and $\Delta m^2_{31} = 2 b (-s)-b^2<0$. In this model, we predict a neutrino mass spectrum with inverted hierarchy. To fit the experimental data, 
we need $b=0.0489$~eV and $s=-3.9 \times 10^{-4}$~eV, which allows to predict the light neutrino mass spectrum explicitly: $|m_1|=0.0486$~eV, $|m_2|=0.0494$~eV, and $|m_3|=0.0004$~eV.

No matter if soft breaking is present or not, the mixing from the neutrino part is in any case given by Eq.~\eqref{eq:bimaximal}, which can be reformulated as
\begin{equation}
 \mathcal{U}^{(1)}_\nu=\left(
\begin{array}{ccc}
 \frac{1}{\sqrt{2}} & \frac{1}{\sqrt{2}} & 0 \\
 -\frac{m_{L1}}{\sqrt{2} \sqrt{m_{L1}^2+m_{L2}^2}} & \frac{m_{L1}}{\sqrt{2}
   \sqrt{m_{L1}^2+m_{L2}^2}} & \frac{m_{L2}}{\sqrt{m_{L1}^2+m_{L2}^2}} \\
 \frac{m_{L2}}{\sqrt{2} \sqrt{m_{L1}^2+m_{L2}^2}} & -\frac{m_{L2}}{\sqrt{2} \sqrt{m_{L1}^2+m_{L2}^2}} & \frac{m_{L1}}{\sqrt{m_{L1}^2+m_{L2}^2}}
\end{array}
\right).
 \label{eq:Unu1}
\end{equation}
Bimaximal mixing is perfectly restored in the limit $m_{L1}=m_{L2}$, which should approximately be true. 
Considering then Eq.~\eqref{eq:UL} for the charged lepton mixing matrix $\mathcal{U}_L$, we will 
then obtain neutrino mixing angles compatible with the experimental values.

\subsection{Case 2 (type II seesaw scenario): $m^{\alpha \beta}_L \ll m^{\alpha i}_D \ll M^{i j}_R$}

Under the hypothesis that $m^{\alpha i}_D \sim m_D$, the eigenvalues of the $6\times 6$ neutrino mass matrix are $\mathcal{E}=\{ \lambda_+, \lambda_-, 0,\Lambda_+, \Lambda_-, 0 \}$, with 
$\lambda_\pm = \pm \sqrt{\frac{2(M^{12}_R-M^{13}_R)^2}{(M^{12}_R)^2+(M^{13}_R)^2}} m_D
+\mathcal{O}(m^3_D)$ 
and 
$\Lambda_\pm=\pm \sqrt{(M^{12}_R)^2+(M^{13}_R)^2} + \mathcal{O}(m^2_D)$. 
As in the previous case, in this scenario we need soft breaking terms to avoid the presence 
of two zero eigenvalues. 
In this case, we have to diagonalize the following mass matrix:\begin{equation}
\mathcal{M}^{(2)}_\nu=
\begin{pmatrix}
\begin{array}{c|c}
\begin{matrix}
s^{e e}_L & m^{e \mu}_L & m^{e \tau}_L \\
m^{e \mu}_L & s^{\mu \mu}_L & 0 \\
m^{e \tau}_L & 0 & s^{\tau \tau}_L 
\end{matrix}
&
\begin{matrix}
m^{e 1}_D & 0 & 0 \\
0 & m^{\mu 2}_D & m^{\mu 3}_D \\
0 & m^{\tau 2}_D & m^{\tau 3}_D
\end{matrix}\\\hline
\begin{matrix}
m^{e 1}_D & 0 & 0 \\
0 & m^{\mu 2}_D & m^{\tau 2}_D \\
0 & m^{\mu 3}_D & m^{\tau 3}_D
\end{matrix}
&
\begin{matrix}
S^{11}_R & M^{12}_R & M^{13}_R \\
M^{12}_R & S^{22}_R & 0 \\
M^{13}_R & 0 & S^{33}_R
\end{matrix}
\end{array}
\end{pmatrix}.
\end{equation}
Defining
\begin{equation}
\mathbf{m_L}\equiv
\begin{pmatrix}
s^{e e}_L & m^{e \mu}_L & m^{e \tau}_L \\
m^{e \mu}_L & s^{\mu \mu}_L & 0 \\
m^{e \tau}_L & 0 & s^{\tau \tau}_L
\end{pmatrix},\:
\mathbf{m_D}\equiv
\begin{pmatrix}
m^{e 1}_D & 0 & 0 \\
0 & m^{\mu 2}_D & m^{\mu 3}_D \\
0 & m^{\tau 2}_D & m^{\tau 3}_D
\end{pmatrix},\:
\mathbf{M_R}\equiv
\begin{pmatrix}
S^{11}_R & M^{12}_R & M^{13}_R \\
M^{12}_R & S^{22}_R & 0 \\
M^{13}_R & 0 & S^{33}_R
\end{pmatrix},
\end{equation}
we can rewrite $\mathcal{M}_\nu^{(2)}$ as:
\begin{equation}
\mathcal{M}_\nu^{(2)}=
\begin{pmatrix}
\mathbf{m_L} & \mathbf{m_D} \\
\mathbf{m_D^T} & \mathbf{M_R}
\end{pmatrix}\,.
\end{equation}
Taking into account that the eigenvalues of the matrices $\mathbf{M_R}$ are 
much bigger than the entries of the matrices $\mathbf{m_L}$ and 
$\mathbf{m_D}$, we can block diagonalize the matrix $\mathcal{M}_\nu$ 
in the following form:
\begin{equation}
\mathcal{M}^{(2),\rm block}_\nu=
\begin{pmatrix}
\mathbf{m_L}-\mathbf{m_D}\mathbf{M^{-1}_R}\mathbf{m^T_D} & 0_{3\times3} \\
0_{3\times3} & \mathbf{M_R}
\end{pmatrix}\,.
\end{equation}
Considering for simplicity $s^{\alpha \alpha}_L \simeq s$, $S^{ii}_R \simeq S$, and $M^{12}_R\simeq M^{13}_R \sim M_R$, and diagonalizing the matrices $\mathbf{m_L}-\mathbf{m_D}\mathbf{M^{-1}_R}\mathbf{m^T_D}$ 
and $\mathbf{M_R}$ separately, we obtain the eigenvalues $\mathcal{E}'=\{ \lambda^\prime_+, \lambda^\prime_-, \lambda_s, \Lambda^\prime_+, \Lambda^\prime_-, \Lambda_s \}$,
with 
$\lambda^\prime_\pm=
s\pm \sqrt{2} \left[ m_L - \frac{m_D^2}{M_R} \right] + \frac{5 m_D^2 S}{4 M_R^2} + 
\mathcal{O}\left(\frac{S^2}{M^3_R}\right)$, 
$\lambda_s=s$, $\Lambda^\prime_\pm=S\pm\sqrt{2} M_R$, and 
$\Lambda_s=S$. Also in this case we could explain the presence of one keV 
sterile neutrino ($S \simeq$ keV) and two heavier sterile neutrinos 
($M_R \gg S$). The light neutrino mass matrix is given by
\begin{equation}
\mathbf{m_L}-\mathbf{m_D}\mathbf{M^{-1}_R}\mathbf{m^T_D}=\left(
\begin{array}{ccc}
 s+\frac{m_D^2 S}{2 M_R^2} & m_L-\frac{m_D^2}{M_R}-\frac{m_D^2 S^2}{2 M_R^3} &
   m_L-\frac{m_D^2}{M_R}-\frac{m_D^2 S^2}{2 M_R^3} \\
 m_L-\frac{m_D^2}{M_R}-\frac{m_D^2 S^2}{2 M_R^3} & s+\frac{m_D^2 S}{M_R^2} &
   \frac{m_D^2 S}{M_R^2} \\
 m_L-\frac{m_D^2}{M_R}-\frac{m_D^2 S^2}{2 M_R^3} & \frac{m_D^2 S}{M_R^2} &
   s+\frac{m_D^2 S}{M_R^2}
\end{array}
\right),
\end{equation}
where we have neglected terms of the order $\mathcal{O}\left(\frac{S^3}{M^4_R}\right)$ and 
where we have assumed that $b\equiv m_L-\frac{m_D^2}{M_R}>0$. 
Similarly to the first case, we predict neutrino masses that are given by $m_1=s+b$, $m_2=s-b$, and 
$m_3=s$. In this case, we again have $b=0.0489$~eV, $s=-3.9 \times 10^{-4}$~eV,  $|m_1|=0.0486$~eV, $|m_2|=0.0494$~eV, and $|m_3|=0.0004$~eV, just as in case 1.

The light neutrino mixing matrix is given by:
\begin{equation}
\mathcal{U}^{(2)}_\nu=\left(
\begin{array}{ccc}
 \frac{1}{\sqrt{2}}-\epsilon  & \frac{1}{\sqrt{2}} +\epsilon & 0 \\
 -\frac{1}{2} - \frac{\epsilon }{\sqrt{2}} & \frac{1}{2}-\frac{\epsilon }{\sqrt{2}} &
   \frac{1}{\sqrt{2}} \\
 \frac{1}{2} + \frac{\epsilon }{\sqrt{2}} & -\frac{1}{2}+\frac{\epsilon }{\sqrt{2}} &
   \frac{1}{\sqrt{2}}
\end{array}
\right),\ {\rm with}\ \eps=\frac{3 m_D^2 S}{16 M_R^2 \left(m_L-\frac{m_D^2}{M_R}\right)}.
\label{eq:mix2}
\end{equation}
Indeed, one can see that without soft breaking ($\eps\equiv 0$), we would end up with a mixing matrix that is exactly given in Eq.~\eqref{eq:bimax}, but the soft breaking is able to alter the form of the matrix. 
It is hence worth to investigate if $\eps$ can be small at all: We know from the condition that the mass spectra have to be compatible with all experiments and observations that $S\simeq \mathcal{O}({\rm keV})\ll M_R$, $m_D\leq \mathcal{O}({\rm 100~GeV})$, and that $(m_L-\frac{m_D^2}{M_R})=\mathcal{O}(m_\nu)\gg s$. Furthermore, remember that we have assumed $m_L-\frac{m_D^2}{M_R}>0$. Using $3/16\sim 0.1$, we have to check if it is possible to have
\begin{equation}
 \left| \frac{m_D^2 S}{M_R^2 \left(m_L-\frac{m_D^2}{M_R}\right)} \right|= \left| \frac{S}{M_R} \times \frac{m_D^2/M_R}{m_L-\frac{m_D^2}{M_R}} \right|\to 0.
 \label{eq:eps-check}
\end{equation}
Here, the first factor $S/M_R$ is always small. The second one has a denominator $m_L-\frac{m_D^2}{M_R}$, which has to be of the order of the neutrino mass. In the limit $m_L\gg \frac{m_D^2}{M_R}$, this factor will be roughly equal to $\frac{m_D^2/M_R}{m_L}$, and hence tiny. Even in the limit $m_L\ll \frac{m_D^2}{M_R}$ it can at most be of order one, which is not enough to compensate the smallness of the first factor. Only in the extremely fine-tuned case, $m_\nu \ll m_L, \frac{m_D^2}{M_R}$, it may be that this factor gets sizable, but this case is not to be expected. 
Considering the expression reported in Eq.~\eqref{eq:UL} for the charged lepton mixing matrix $\mathcal{U}_L$, we find neutrino mixing angles compatible with the experimental values. 
Accordingly, this model provides a natural benchmark scenario for keV neutrino Dark Matter that even works without additional requirements such as the one needed for case 1.

\subsection{Case 3 (hybrid scenario): $m^{\alpha \beta}_L \sim m^{\alpha i}_D \ll M^{i j}_R$}

Under the hypotheses that $m^{\alpha i}_D \sim m^{\alpha \beta}_L \sim m_L$ and $M^{i j}_R \sim M_R$, 
the eigenvalues of the $6\times 6$ neutrino mass matrix are given by $\mathcal{E}=\{ \lambda_+, \lambda_-, 0, \Lambda_+, \Lambda_-, 0 \}$, 
with 
$\lambda_\pm= \pm \left(\sqrt{2} m_L-\sqrt{2}\frac{m^2_L}{M_R}\right)
+\mathcal{O}\left(\frac{m^3_L}{M^2_R}\right)$ and 
$\Lambda_\pm =\pm \left(\sqrt{2}M_R+\frac{5 m^2_L}{2 \sqrt{2}M_R}\right)
+\mathcal{O}\left(\frac{m^3_L}{M^2_R}\right)$. 
As in the previous cases, we need soft breaking terms to avoid the presence 
of two zero eigenvalues. In this case, we have to diagonalize the following mass matrix:
\begin{equation}
\mathcal{M}^{(3)}_\nu=
\begin{pmatrix}
\begin{array}{c|c}
\begin{matrix}
s^{e e}_L & m^{e \mu}_L & m^{e \tau}_L \\
m^{e \mu}_L & s^{\mu \mu}_L & 0 \\
m^{e \tau}_L & 0 & s^{\tau \tau}_L 
\end{matrix}
&
\begin{matrix}
m^{e 1}_D & 0 & 0 \\
0 & m^{\mu 2}_D & m^{\mu 3}_D \\
0 & m^{\tau 2}_D & m^{\tau 3}_D
\end{matrix}\\\hline
\begin{matrix}
m^{e 1}_D & 0 & 0 \\
0 & m^{\mu 2}_D & m^{\tau 2}_D \\
0 & m^{\mu 3}_D & m^{\tau 3}_D
\end{matrix}
&
\begin{matrix}
S^{11}_R & M^{12}_R & M^{13}_R \\
M^{12}_R & S^{22}_R & 0 \\
M^{13}_R & 0 & S^{33}_R
\end{matrix}
\end{array}
\end{pmatrix}.
\end{equation}
Using the same approximations as for the scenario in case 2, we obtain the eigenvalues $\mathcal{E}=\{ \lambda^\prime_+, \lambda^\prime_-, \lambda_s, 
\Lambda^\prime_+, \Lambda^\prime_-, \Lambda_s \}$, 
with $\lambda^\prime_\pm=
s\pm \sqrt{2} m_L \left(1- \frac{m_L}{M_R} \right)+\mathcal{O}\left(\frac{m^2_L}{M^2_R}\right)$, $\lambda_s=s$, $\Lambda^\prime_\pm=S\pm\sqrt{2} M_R$, and $\Lambda_s=S$. Similar to before we have $b=0.0489$~eV, $s=-3.9 \times 10^{-4}$~eV,  $|m_1|=0.0486$~eV, $|m_2|=0.0494$~eV, and $|m_3|=0.0004$~eV, where $b=\sqrt{2} m_L \left(1- \frac{m_L}{M_R} \right)$. Note that (due to $m_L\ll M_R$) $b>0$ is the only possibility in this case.

The mixing matrix for light neutrinos is given by
\begin{equation}
 \mathcal{U}^{(3)}_\nu= \left(
\begin{array}{ccc}
 \frac{1}{\sqrt{2}}-\frac{\eps}{4} & \frac{1}{\sqrt{2}}+\frac{\eps}{4} & 0 \\
 -\frac{1}{2} - \frac{\eps}{4 \sqrt{2}} & \frac{1}{2}-\frac{\eps}{4
   \sqrt{2}} & \frac{1}{\sqrt{2}} \\
 \frac{1}{2} + \frac{\eps}{4 \sqrt{2}} & -\frac{1}{2}+\frac{\eps}{4\sqrt{2}} & \frac{1}{\sqrt{2}}
\end{array}
\right),\ {\rm with}\ \eps=\frac{3 m_L S}{4 M_R(M_R-m_L)}\simeq \frac{3 m_L S}{4 M^2_R}\,.
 \label{eq:mix3}
\end{equation}
In this case, $\eps$ will always be tiny and could even be taken to be zero. 
Therefore, also this model will lead to neutrino mixing angles compatible with the experimental values.

\section{\label{sec:conclusions} Conclusions}

We have studied how a  $L_e-L_\mu-L_\tau$ flavour symmetry can be used to simultaneously explain observed masses and mixings of the light (active) neutrinos and how it simultaneously can lead to a heavy (sterile) neutrino sector, where one state is light. Such a flavour symmetry would then be responsible for a scenario with keV sterile neutrinos, which are a prime candidate for Warm Dark Matter. 
The $L_e-L_\mu-L_\tau$ flavour symmetry chosen actually predicts two neutrinos with nearly the same mass and one which is massless, in the sterile as well as in the active sector. Soft symmetry breaking, however, creates a non-zero mass for these particles, which will be much smaller than the masses of the respective degenerate pairs. Furthermore, the symmetry predicts zero $\theta_{13}$ and maximal $\theta_{23}$ and $\theta_{12}$, where the last prediction has to be corrected by the mixing coming from the charged lepton sector in order to lead to the leptonic mixing we expect to be present.

\section*{Acknowledgements}

We are grateful to Werner Rodejohann for fruitful discussions and important suggestions. This work has been supported by the DFG Sonderforschungsbereich Transregio 27 Neutrinos and beyond Weakly interacting particles in Physics, Astrophysics and Cosmology. The work of AM is supported by the Royal Institute of Technology (KTH), under project no. SII-56510.

\newcommand{\eprint}[1]{
arXiv: \href{http://arxiv.org/abs/#1}{\texttt{#1}}
}

\bibliographystyle{JHEP}
\bibliography{keVneutrino}

\end{document}